\documentclass[10pt, a4paper]{article}
\usepackage{supertabular, setspace,hyperref}
\usepackage{amsmath,amssymb,amsthm,textcomp}
\usepackage{amsfonts,graphicx}
\usepackage[mathscr]{eucal}
\usepackage{color}
\usepackage{float}
\usepackage[caption = false]{subfig}
\usepackage{tikz}
\usetikzlibrary
{shapes,arrows,chains,matrix,positioning,scopes,shadows,calc,snakes}
\theoremstyle{definition}

\numberwithin{equation}{section}

\newcommand{\ncom}{\newcommand}

\ncom{\beq}{\begin{equation}}
\ncom{\eeq}{\end{equation}}
\ncom{\bea}{\begin{eqnarray*}}
\ncom{\eea}{\end{eqnarray*}}
\ncom{\beqa}{\begin{eqnarray}}
\ncom{\eeqa}{\end{eqnarray}}
\ncom{\nno}{\nonumber}
\ncom{\non}{\nonumber}
\ncom{\ds}{\displaystyle}
\ncom{\half}{\frac{1}{2}}
\ncom{\mbx}{\makebox{.25cm}}
\ncom{\hs}{\mbox{\hspace{.25cm}}}
\ncom{\rar}{\rightarrow}
\ncom{\Rar}{\Rightarrow}
\ncom{\noin}{\noindent}
\ncom{\bc}{\begin{center}}
\ncom{\ec}{\end{center}}
\ncom{\sz}{\scriptsize}
\ncom{\rf}{\ref}
\ncom{\s}{\sqrt{2}}
\ncom{\sgm}{\sigma}
\ncom{\Sgm}{\Sigma}
\ncom{\psgm}{\sigma^{\prime}}
\ncom{\dt}{\delta}
\ncom{\Dt}{\Delta}
\ncom{\lmd}{\lambda}
\ncom{\Lmd}{\Lambda}
\ncom{\Th}{\Theta}
\ncom{\e}{\eta}
\ncom{\eps}{\epsilon}
\ncom{\pcc}{\stackrel{P}{>}}
\ncom{\lp}{\stackrel{L_{p}}{>}}
\ncom{\dist}{{\rm\,dist}}
\ncom{\sspan}{{\rm\,span}}
\ncom{\re}{{\rm Re\,}}
\ncom{\im}{{\rm Im\,}}
\ncom{\sgn}{{\rm sgn\,}}
\ncom{\ba}{\begin{array}}
\ncom{\ea}{\end{array}}
\ncom{\hone}{\mbox{\hspace{1em}}}
\ncom{\htwo}{\mbox{\hspace{2em}}}
\ncom{\hthree}{\mbox{\hspace{3em}}}
\ncom{\hfour}{\mbox{\hspace{4em}}}
\ncom{\vone}{\vskip 2ex}
\ncom{\vtwo}{\vskip 4ex}
\ncom{\vonee}{\vskip 1.5ex}
\ncom{\vthree}{\vskip 6ex}
\ncom{\vfour}{\vspace*{8ex}}
\ncom{\norm}{\|\;\;\|}
\ncom{\integ}[4]{\int_{#1}^{#2}\,{#3}\,d{#4}}
\ncom{\vspan}[1]{{{\rm\,span}\{ #1 \}}}
\ncom{\dm}[1]{ {\displaystyle{#1} } }
\ncom{\ri}[1]{{#1} \index{#1}}

\newtheoremstyle
   {remarkstyle}
   {}
   {11pt}
   {}
   {}
   {\bfseries}
   {:}
   {     }
   {\thmname{#1} \thmnumber{#2} }

\theoremstyle{remarkstyle}

\def\eps{\varepsilon}

\begin{document}

\title{On weak shock diffraction in real gases}

%\shorttitle{On weak shock diffraction in real gases} %%%for recto running head
%\shortauthorlist{Neelam Gupta and V. D. Sharma} %%% for verso running head
\author{\bf Neelam Gupta and V. D. Sharma\\
{\it Department of Mathematics,
Indian Institute of Technology Bombay,}\\
{\it Mumbai-400076}}

%%%%%%%
%\and
%%%%%%% Third author details
%\name{Insert third author}
%\address{Third author address}}

\maketitle

\begin{abstract}
Asymptotic solutions are obtained for the two-dimensional Euler system for real gases with appropriate boundary conditions which describe the diffraction of a weak shock at a right-angled wedge; the real gas effects are characterized by a van der Waals type equation of state. The behavior of the flow configuration influenced by the real gas effects, that includes the local structure near a singular point, is studied in detail. 
\end{abstract}
\section{Introduction}
Shock waves were recognized as a natural phenomenon more than a century ago, yet they are still not widely understood. The problem of shock reflection-diffraction by wedges has a salient feature in gasdynamics which has captured the interest of researchers over the last fifty  years (see, Courant \& Friedrichs \cite{courant1976}, Glass \& Sisilian \cite{glass}, and Ben-dor \cite{dor}). When a weak incident shock reflects off a wedge, various configurations are generated including the regular and Mach reflections (see, Chang \& Chen \cite{chen1986}, Zheng \cite{zheng,zheng1}, Chen \cite{chen}, and Chang \& Hsiao \cite{chang}). Here, we consider the case when regular reflection takes place for sufficiently large wedge angles. When a plane shock hits a wedge head on; it is reflected by the wedge surface, and at the same time the flow behind it is diffracted by the compressive corner of the wedge and produces circular waves in the vicinity of the corner. Moreover, the circular diffracted waves propagate with the sound speed in the medium surrounded by the wedge. This remarkable configuration so-called reflection-diffraction phenomena has been studied previously, within the context of an ideal gas, by Keller \& Blank \cite{keller1951}, Hunter \& Keller \cite{kel}, Harabetian \cite{eduard1987}, Morawetz \cite{morawetz}, Zheng \cite{zheng2}, Hunter \& Brio \cite{Hunter2000}, Canic et al. \cite{keyfitz}, Rosales \& Tabak \cite{tabak}, and Hunter \& Tesdall \cite{tesdall, tesdall2}. The ideal gas law is based on the assumption that gases are composed of point masses that undergo perfectly elastic collisions; but at a low temperature or high pressure, behavior of gases deviates from the ideal gas law and follows van der Waals type gas that deals with the possible real gas effects \cite{ Wu1996, arora, manoj, manoj1}. In this paper, we use an asymptotic approach to the shock diffraction problem when the real gas effects are taken into account; the real gas effects, presented here, are characterized by a van der Waals type equation of state. \\
The set up of the reflection consists of a straight shock hitting a right-angled wedge at the origin at time $t=0$; the shock is assumed to be weak moving parallel to one side of the wedge, which is placed parallel to the y-axis (see Figure 1). The shock is then reflected and diffracted off the wedge; the linearized solution to this problem was presented in \cite{keller1951}; the solution was modified later in \cite{kel} using the theory of weakly nonlinear geometrical acoustics. Here, we use asymptotic expansions to obtain nonlinear corrections to the behavior of diffracted wave near the wavefront; in the limit of vanishing van der Waals excluded volume, we recover the results obtained in \cite{kel}. The main objective of the present paper is to study how the real gas effects influence the behavior of the reflected and diffracted wavefronts, and in particular the local structure of the self-similar solutions of the Euler equations near a singular point; the motivation stems from the work carried out in \cite{keller1951, kel, eduard1987, myers, hunter88}.\\
This paper is structured as follows: In section $2$, we set up the mathematical model with appropriate boundary conditions to describe the shock reflection-diffraction phenomena. In section $3$, we provide asymptotic expansions of the Rankine-Hugoniot conditions for incident and reflected shocks and obtain a piecewise leading order constant solution to $O(\epsilon)$ in the exterior region. In section $4$, we obtain first order approximation to the problem in the diffracted wave region; this solution is not valid near the boundary points where the governing system becomes degenerate. Indeed, the solution has a singularity of the type $\sqrt{\zeta-a_0}$ at these points. So, there is a need to find a different expansion near such points. In order to achieve this objective, we rescale the independent variables near these points and find a new approximation in section $5$. In a close neighborhood of the point $Q(a_0, \pi)$ (see Figure 1), since the change in tangential direction is faster than the radial direction, we present another expansion in section $6$ and obtain an asymptotic solution in the neighborhood of the singular point. Finally, we match all the approximations to boundary conditions in order to get a uniformly valid solution throughout the flow field. We summarize our conclusions in section $7$.
\section{Shock reflection-diffraction configuration}
Consider the compressible Euler equations of gasdynamics in two-dimensional space
%%%%%%%%%%%%%%%%%%%%%%%%%%%%%%%%%%%%%%%%%%%%%%%%%%%%%%%%%%%%%%%%%%%%%%%%%%%%%%%%%%%%%%%%%%%%%%%%%%%%%%%%%
\begin{align}\label{equ1}
\begin{split}
\rho_{t} + ({\rho}{u})_x + ({\rho}{v})_y ={0},\\  
({\rho}{u})_{t} + ({\rho}{u^2}+p)_x + ({\rho}{u}{v})_y ={0},\\
({\rho}{v})_{t} + ({\rho}{u}{v})_x + ({\rho}{v^2}+p)_y ={0},\\
\left({\rho}{\Big(e+\frac{{u^2}+{v^2}}{2}\Big)}\right)_{t} + \left({\rho}{u}{\Big(h+\frac{{u^2}+{v^2}}{2}\Big)}\right)_{x}+\left({\rho}{v}{\Big(h+\frac{{u^2}+{v^2}}{2}\Big)}\right)_{y}=0,
\end{split}
\end{align}
for the variables $(\rho, u, v, p, e)$, where $\rho$, ($u$, $v$), $p$, $e$, and $h$ are the density, velocity components, pressure, internal energy, and specific enthalpy, respectively. In addition, $e$ and $h$, the functions of $\rho$ and $p$, are related by the second law of  thermodynamics
\begin{equation*}
 Td{\mbox{\scriptsize{S}}}=de+pdV=dh-Vdp,
\end{equation*}
with $T(\rho, p)$ being the temperature, $V$ the specific volume and ${\mbox{\scriptsize{S}}}(\rho, p)$ the specific entropy. Here, we consider the fact that the gas obeys a van der Waals type equation of state, for which pressure $p$, temperature $T$,  specific volume $V$, internal energy $e$, entropy $\mbox{\scriptsize{S}}$, and specific enthalpy $h$ are related as 
\begin{equation}\label{equ}
p=\frac{RT}{(V-b)},~~~e=\frac{p(V-b)}{\gamma-1},~~~{\mbox{\scriptsize{S}}}=c_v\ln{(p(V-b)^{\gamma})}+constant,~~~h=\frac{p(\gamma{V}-b)}{\gamma-1},
\end{equation} 
where R is the gas constant, $\gamma(>1)$ the ratio of specific heats, and $b$ the van der Waals excluded volume. \\
When a weak shock in the $(x, y, t)$ coordinates, with state ahead of the shock ${(\rho, u, v, p)}={(\rho_0, 0, 0,  p_0)}$ for some $p_0>0$ and the state behind the shock ${( \rho_1, u_1, 0, p_1)}$ with $p_1>p_0$, hits a right-angled wedge head on, it reproduces a diffraction-reflection phenomenon. Therefore, we seek a solution of the system (\ref{equ1}) satisfying the initial condition 
\begin{equation}\label{equ2}
{(\rho, u, v, p)}\Big|_{t=0} = \left\{
  \begin{array}{l l}
    {(\rho_0, 0, 0,  p_0)}, & y> 0,~ x>0\\
    {( \rho_1, u_1, 0, p_1)}, & x<0,
\end{array} \right.
\end{equation}
and the slip boundary condition along the wedge
\begin{equation}\label{equ3}
v={0}\Big|_{y=0}\hspace{1cm} x>0,\hspace{.2cm} t>0.
\end{equation}
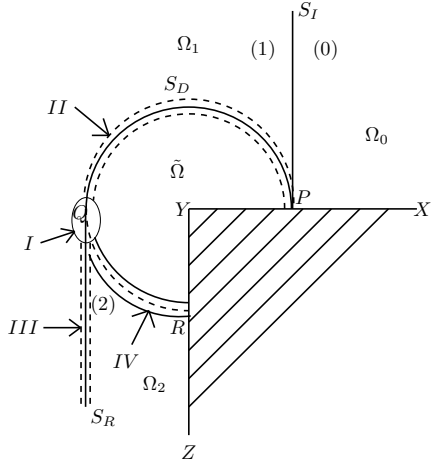
\begin{figure}
\centering
\scalebox{.75}{
			\begin{tikzpicture}
			\draw (-1.8, 6.3) ellipse (.25cm and .4cm);
			\begin{scope}
			\draw [thick](0,2.5)--(0,6.5);
			\draw [thick](0,6.5)--(4,6.5);
			\draw [thick](1.8, 6.5) arc (0: 180: 1.8);
			\draw [thick][dashed](-1.8, 6.5) arc (180: 270: 1.8);
			\draw [thick][dashed](1.84, 6.6) arc (0: 175: 1.84);
			\draw [thick][dashed](1.68, 6.5) arc (0: 175: 1.68);
			\draw [thick](-1.65, 6) arc (200: 270: 1.76);
			\draw [thick](-1.75, 5.7) arc (200: 277: 1.68);
			\draw [thick](1.82,6.5)--(1.82, 10);
			\draw [thick][dashed](-1.89,5.9)--(-1.89, 3);
			\draw [thick][dashed](-1.73,5.9)--(-1.73, 3);
			\draw [thick](-1.81,6.5)--(-1.81, 3);
			\draw [thick](.5,6.5)--(0, 6);
			\draw [thick](1,6.5)--(0, 5.5);
			\draw [thick](1.5,6.5)--(0, 5);
			\draw [thick](2,6.5)--(0, 4.5);
			\draw [thick](2.5,6.5)--(0, 4);
			\draw [thick](3,6.5)--(0, 3.5);
			\draw [thick](3.5,6.5)--(0, 3);
			\draw [thick](-2.6, 5.9)--(-2,6.1);
			\draw [thick](-2, 6.1)--(-2.15,5.9);
			\draw [thick](-2, 6.1)--(-2.25,6.15);
			\draw [thick](-1.4, 7.8)--(-2,8.3);
			\draw [thick](-1.4, 7.8)--(-1.7,7.8);
			\draw [thick](-1.4, 7.8)--(-1.5,8.1);
			\draw [thick](-1.9,4.4)--(-2.55, 4.4);
			\draw [thick](-1.9, 4.4)--(-2.1,4.25);
			\draw [thick](-1.9, 4.4)--(-2.1,4.55);
			\draw [thick](-.7,4.7)--(-1, 4.1);
			\draw [thick](-.7,4.7)--(-1, 4.6);
			\draw [thick](-.7,4.7)--(-.6, 4.4);
			\end{scope}
			\draw (2.4,9.3) node {$(0)$};
			\draw (1.3,9.3) node {$(1)$};
			\draw (-1.5,4.8) node {$(2)$};
			\draw (2,6.7) node {$P$};
			\draw (-1.9,6.4) node {$Q$};
			\draw (-.2,4.4) node {$R$};
			\draw (2.1,10) node {$S_I$};
			\draw (-1.5,2.8) node {$S_R$};
			\draw (-.2,8.7) node {$S_D$};
			\draw (4.1,6.5) node {$X$};
			\draw (-0.1,6.5) node {$Y$};
			\draw (0,2.2) node {$Z$};
			%\draw (4,9) node {$I$};
			%\draw (-1,9.5) node {$I$};
			%\draw (-.7,6.5) node {$I$};
			\draw (-2.8,5.9) node {$I$};
			\draw (-2.3,8.3) node {$II$};
			\draw (-2.9,4.4) node {$III$};
			\draw (-1.1,3.8) node {$IV$};
			\draw (3.3,7.8) node {$\Omega_0$};
			\draw (0,9.4) node {$\Omega_1$};
			\draw (-.6,3.4) node {$\Omega_2$};
			\draw (-.2,7.2) node {$\tilde{\Omega}$};
\end{tikzpicture}
}
\caption{\textit{Self-similar flow pattern at a right-angled wedge.}}
\label{figure1}
\end{figure}
%%%%%%%%%%%%%%%%%%%%%%%%%%%%%%%%%%%%%%%%%%%%%%%%%%%%%%%%%%%%%%%%%%%%%%%%%%%%%%%%%%%%%%%%%%%%%%%
%\section{Self-similar flow and shock reflection-diffraction configuration}
It may be observed that the equations (\ref{equ1}), governing the diffraction-reflection phenomenon, together with initial and boundary conditions (\ref{equ2}) and (\ref{equ3}) are invariant under the self-similar scaling:
\begin{equation}\nonumber
(t, x, y)\rightarrow (\nu{t}, \nu{x}, \nu{y}),~~~~~~{(\rho, u, v, p)}(t, x, y)={(\rho, u, v, p)}(\nu{t}, \nu{x}, \nu{y}),
\end{equation}
where $\nu>0$ is an arbitrary constant. Thus, introducing the variables ~$\zeta=\frac{\sqrt{{x^2}+{y^2}}}{t}$~ and $\beta=tan^{-1}\left(\frac{y}{x}\right)$, the Euler equations (\ref{equ1}) for self-similar flow consist of the conservation law of mass, momentum, and energy: 
\begin{align}\label{equ5}
\begin{split}
&{(\mbox{\scriptsize{U}}-\zeta)}{\rho}_{\zeta}+\rho{\mbox{\scriptsize{U}}}_{\zeta}+{1}/{\zeta}(\rho{\mbox{\scriptsize{V}}}_{\beta}+\mbox{\scriptsize{V}}{\rho}_{\beta}+\rho\mbox{\scriptsize{U}})=0,\\&
({\mbox{\scriptsize{U}}-\zeta)}{\mbox{\scriptsize{U}}}_{\zeta}+(1/{\rho})p_{\zeta}+({\mbox{\scriptsize{V}}}/{\zeta}){\mbox{\scriptsize{U}}}_{\beta}-{{\mbox{\scriptsize{V}}}^2}/{\zeta}=0,\\&
({\mbox{\scriptsize{U}}}-\zeta){\mbox{\scriptsize{V}}}_{\zeta}+({\mbox{\scriptsize{V}}}/{\zeta}){\mbox{\scriptsize{V}}}_{\beta}+(1/{\rho})p_{\beta}+{\mbox{\scriptsize{U}}}{\mbox{\scriptsize{V}}}/{\zeta}=0,\\&
-{\zeta}\left({\rho}{\Big(e+\frac{{{\mbox{\scriptsize{U}}}^2}+{{\mbox{\scriptsize{V}}}^2}}{2}\Big)}\right)_{\zeta}+\left({\rho}{\mbox{\scriptsize{U}}}{\Big(h+\frac{{{\mbox{\scriptsize{U}}}^2}+{{\mbox{\scriptsize{V}}}^2}}{2}\Big)}\right)_{\zeta}+(1/{\zeta})\left({\rho}{\mbox{\scriptsize{V}}}{\Big(h+\frac{{{\mbox{\scriptsize{U}}}^2}+{{\mbox{\scriptsize{V}}}^2}}{2}\Big)}\right)_{\beta}+({\rho \mbox{\scriptsize{U}}}/{\zeta}){\Big(h+\frac{{{\mbox{\scriptsize{U}}}^2}+{{\mbox{\scriptsize{V}}}^2}}{2}\Big)}=0,
\end{split}
\end{align}
with initial and boundary conditions:
\begin{equation}\label{equ6}
 \lim_{{\zeta} \to \infty}{(\rho, {\mbox{\scriptsize{U}}}, {\mbox{\scriptsize{V}}}, {p})}
  = \left\{
  \begin{array}{l l}
    {(\rho_0, 0, 0, p_0)}, & 0 \leq{\beta}<{\frac{\pi}{2}},\\
		{(\rho_1, u_1\cos{\beta}, -u_1\sin{\beta}, p_1)}, & {\frac{\pi}{2}}\leq{\beta}<{\frac{3\pi}{2}},
\end{array} \right.
\end{equation}
and 
\begin{equation}\label{equ7}
\mbox{\scriptsize{V}}={0}\Big|_{\beta=0},
\end{equation}
where $h$ is given by $(\ref{equ})_4$, $\beta=\theta-\pi/4$, with $\beta=0$ on the edge $XY$,  
\begin{equation}\label{equ5*}
\mbox{\scriptsize{U}}=u\cos{\beta}+v\sin{\beta},~~ \mbox{\scriptsize{V}}=-u\sin{\beta}+v\cos{\beta}, 
\end{equation} 
and the sound speed is given by~~$a(\rho, {\mbox{\scriptsize{S}}})=\sqrt{\partial{p}/\partial{\rho}}=\sqrt{\dfrac{{\gamma}p}{\rho(1-b\rho)}}$,~~~~$0\leq{b\rho}<1$.\\
 For smooth solutions, the energy equation $(\ref{equ5})_4$ may be written as 
\begin{equation}\label{equS}
(\mbox{\scriptsize{U}}-\zeta)\mbox{\scriptsize{S}}_{\zeta}+({{\mbox{\scriptsize{V}}}/{\zeta}})\mbox{\scriptsize{S}}_{\beta}=0.
\end{equation} 
In view of $(\ref{equ})_3$, equations $(\ref{equ5})_{1,2,3}$ and (\ref{equS}) can be written using vector matrix notation as
\begin{equation}\label{equ8}
{(A(W)-{\zeta}I)}W_{\zeta}+(1/{\zeta})B(W)W_{\beta}+(1/{\zeta})C(W)W=0,
\end{equation}
where $W=(\rho, \mbox{\scriptsize{U}}, \mbox{\scriptsize{V}}, \mbox{\scriptsize{S}})^T$ and $A=(A_{ij})$, $B=(B_{ij})$, and $C=(C_{ij})$ are $4\times4$ matrices with nonzero entries $A_{11}=\mbox{\scriptsize{U}}$, $A_{12}=\rho$, $A_{21}={a^2}/{\rho}$, $A_{22}=\mbox{\scriptsize{U}}$, $A_{24}={p}/{\rho{c_v}}$, $A_{33}=\mbox{\scriptsize{U}}$, $A_{44}=\mbox{\scriptsize{U}}$, $B_{11}=\mbox{\scriptsize{V}}$, $B_{13}=\rho$, $B_{22}=\mbox{\scriptsize{V}}$, $B_{31}={a^2}/{\rho}$, $B_{33}=\mbox{\scriptsize{V}}$, $B_{34}={p}/{\rho{c_v}}$, $B_{44}=\mbox{\scriptsize{V}}$, $C_{12}=\rho$, $C_{23}=-\mbox{\scriptsize{V}}$, and $C_{32}=\mbox{\scriptsize{V}}$; the remaining entries are all zero.\\
Initially at time $t=0$, it is assumed that the shock front hits the right-angled wedge head on. At time $t>0$, it propagates further along the wedge, a part of it is diffracted by the vertex and produces a nearly circular diffracted wave originating from the vertex of the wedge moving at sonic speed, whereas other part is reflected back by the wedge (see Figure $1$). The unknown curved portion $PQ$, due to the influence of the vertex $Y$, joins the diffracted wavefront $QR$ at $Q(a_0, \pi)$ and gives rise the overall shock diffraction-reflection phenomenon. The location of the incident shock after it has moved beyond the domain of the influence of the origin (vertex of the wedge) is given by
\begin{equation}\label{equ7**}
S_I:~~\zeta=a_0\sec{\beta},
\end{equation}
where $a_0=\sqrt{{{\gamma}p_0}/{\rho_0(1-b\rho_0)}}$.\\
 Similarly, the equation of the reflected shock  is given by 
\begin{equation}\label{equ7*}
S_R:~~\zeta=-{a_0}\sec{\beta}.
\end{equation}
It follows from (\ref{equ7**}) and (\ref{equ7*}) that ${d{\zeta}}/{d\tilde{b}}>0$ for incident shock $S_I$ while for reflected shock $S_R$ ${d{\zeta}}/{d\tilde{b}}<0$, where $\tilde{b}=b\rho_0$; this implies the domain of linearized solution becomes larger with an increase in $\tilde{b}$.\\
It may be noticed that the domain of entire flow field consists of the following regions:
 \begin{align}\nonumber
\begin{split}
&\Omega_0=\left\{(\zeta, \beta): \zeta>{a_0}\sec{\beta}\right\},\\&
\Omega_1=\left\{(\zeta, \beta): a_0<\zeta<{a_0}\sec{\beta},~~ 0<\beta<\pi\right\}\cup \left\{(\zeta, \beta): \zeta<-{a_0}\sec{\beta},~~ \pi<\beta<{{3\pi}/2}\right\},\\&
\Omega_2=\left\{(\zeta, \beta): -{a_0}\sec{\beta}<\zeta<-{a_0},~~ \pi<\beta<{{3\pi}/2}\right\},~~\widetilde{\Omega}=\left\{(\zeta, \beta): \zeta<{a_0},~~ 0<\beta<{{3\pi}/2}\right\}.
\end{split}
\end{align}
Here $QR$ is referred to as the sonic curve, $\zeta=a_0$, across which there is a continuous transition from the supersonic region $\Omega_2$ to the subsonic region $\tilde{\Omega}$, whereas $PQ$ is the free boundary, called diffraction of the planar shock, across which the transition undergoes a jump from the supersonic region ${\Omega}_1$ to the subsonic region $\tilde{\Omega}$ near the origin. The flow is constant in the regions $\Omega_1$ and $\Omega_2$ while it is pseudo-subsonic in $\tilde{\Omega}$. 
The nature of this flow pattern is regulated by Euler system (\ref{equ1}), which is hyperbolic; however, in self-similar coordinates, the corresponding flow is governed by mixed type equations (\ref{equ8}). Indeed, the system (\ref{equ8}) changes its type from elliptic to hyperbolic when the point $(\zeta, \beta)$ runs from the origin to infinity.
Hence in order to determine the entire flow field and the wave structure, one needs to solve the free boundary value problem for a degenerate elliptic equation.
The system (\ref{equ8}) has four real eigenvalues 
\begin{align}\label{equ10}
\begin{split}
&\lambda=\frac{\mbox{\scriptsize{V}}}{\zeta(\mbox{\scriptsize{U}}-\zeta)},~~~\text{(multiplicity-2)}\\&
\lambda=\frac{\mbox{\scriptsize{V}}(\mbox{\scriptsize{U}}-\zeta){\pm}a{\sqrt{\mbox{\scriptsize{V}}^2-a^2+(\mbox{\scriptsize{U}}-\zeta)^2}}}{{\zeta}\left((\mbox{\scriptsize{U}}-\zeta)^2-a^2\right)},
\end{split}
\end{align}
with $\mbox{\scriptsize{V}}^2+(\mbox{\scriptsize{U}}-\zeta)^2>a^2$. Therefore, equations (\ref{equ10}) imply that the system (\ref{equ8}) is hyperbolic and the flow is supersonic. When $\mbox{\scriptsize{V}}^2+(\mbox{\scriptsize{U}}-\zeta)^2<a^2$, the system is mixed type as two equations in (\ref{equ8}) are hyperbolic and the other two are elliptic. However, $\mbox{\scriptsize{V}}^2+(\mbox{\scriptsize{U}}-\zeta)^2=a^2$ represents a sonic curve in $(\zeta, \beta)$ plane. In general, the system (\ref{equ8}) is mixed type and the flow is transonic.
%%%%%%%%%%%%%%%%%%%%%%%%%%%%%%%%%%%%%%%%%%%%%%%%%%%%%%%%%%%%%%%%%%%%%%%%%%%%%%%%%%%%%%%%%%%%%%%%%%%%%%%%%%%%%
\section{Rankine-Hugoniot conditions for the incident and reflected shocks}
The state ahead of the incident shock is  $(\rho_0, 0, 0, p_0)$; however, in order to find the states $(\rho_1, {\mbox{\scriptsize{U}}}_1, {\mbox{\scriptsize{V}}}_1, p_1)$ and  $(\rho_2, {\mbox{\scriptsize{U}}}_2, {\mbox{\scriptsize{V}}}_2, p_2)$, immediately behind the incident and reflected shocks  denoted by subscripts-1 and -2, respectively, we need the Rankine-Hugoniot (R-H) conditions in 2D. Let $\zeta=G(\beta)$ be the shock curve; then from (\ref{equ1}) Rankine-Hugoniot relations imply
\begin{align}\label{equ11}
\begin{split}
&{G'}[{\rho}]={\mu}[\rho{u}]+{\nu}[{\rho{v}}],\\&
{G'}[{\rho}u]={\mu}[\rho{u^2}+p]+{\nu}[{\rho}{u}{v}],\\&
{G'}[{\rho}v]={\mu}[{\rho}{u}{v}]+{\nu}[\rho{v^2}+p],\\&
{G'}\left[{\rho}{\Big(e+\frac{{u^2}+{v^2}}{2}\Big)}\right]={\mu}\left[\rho u{\Big(h+\frac{{u^2}+{v^2}}{2}\Big)}\right]+{\nu}\left[\rho v{\Big(h+\frac{{u^2}+{v^2}}{2}\Big)}\right],
\end{split}
\end{align} 
where $G'$, $(\mu, \nu)$, and square brackets, [.], denote the shock speed, normal vector to the shock front,  and jump across the shock, respectively.\\
Then using (\ref{equ}), (\ref{equ7**}), and (\ref{equ2}) in (\ref{equ11}), the R-H conditions (\ref{equ11}) on the incident shock give the following relations
\begin{align}\label{equ8*}
\begin{split}
&\frac{p_{1}}{p_0}=\frac{(\gamma+1)\rho_{1}-(\gamma-1)\rho_0-2\tilde{b}{\rho_{1}}}{(\gamma+1)\rho_0-(\gamma-1)\rho_{1}-2\tilde{b}{\rho_{1}}},~~~~~~~~~\rho_1>\rho_0\\&
u_1=\left(\frac{(p_1-p_0)(\rho_1-\rho_0)}{{\rho_0}{\rho}_{1}}\right)^{1/2},~~~\zeta=a_0\sec{\beta},~~~~v_1=0.
\end{split}
\end{align}
Let $\epsilon>0$ be a dimensionless parameter measuring the shock strength, i.e.,
\begin{equation}\label{equ41}
\epsilon=(\rho_1-\rho_0)/{\rho_0}.
\end{equation}
Then in view of (\ref{equ5*}) and (\ref{equ41}), equations (\ref{equ8*}) yield the following asymptotic expansions  of the state-1 variables as $\epsilon\rightarrow0$:
\begin{align}\label{equ42}
\begin{split}
&\frac{\rho_1}{\rho_0}=1+{\rho_{1}^{(1)}}\epsilon,~~~~~~~~~~~~~~~~~~~~~~~~~~~~~~~~~~~~~~\frac{p_1}{p_0}=1+{p_{1}^{(1)}}\epsilon+p_{1}^{(2)}\epsilon^2+O(\epsilon^3),\\&
\frac{\mbox{\scriptsize{U}}_1}{c_0}={\mbox{\scriptsize{U}}_{1}^{(1)}}\epsilon+{\mbox{\scriptsize{U}}_{1}^{(2)}}{\epsilon}^2+O(\epsilon^3),~~~~~~~~~~~~~~~~~\frac{\mbox{\scriptsize{V}}_1}{c_0}={\mbox{\scriptsize{V}}_{1}^{(1)}}\epsilon+{\mbox{\scriptsize{V}}_{1}^{(2)}}{\epsilon}^2+O(\epsilon^3),\\&
\frac{a_1}{c_0}=\kappa_0+{a_{1}^{(1)}}\epsilon+{a_{1}^{(2)}}\epsilon^2+O(\epsilon^3), ~~~~~~~~~\frac{\mbox{\scriptsize{S}}_{1}-\mbox{\scriptsize{S}}_{0}}{c_v}=\frac{\gamma\epsilon^3}{12(1-\tilde{b})^3}(\gamma^2-1)+O(\epsilon^4),\\&
\frac{\zeta}{c_0}={\kappa_0}\sec{\beta}+\frac{{\kappa_0}(\gamma+1)\epsilon}{4(1-\tilde{b})}\sec{\beta}+O(\epsilon^2),
\end{split}
\end{align}
where $\rho_{1}^{(1)}=1$, $p_{1}^{(1)}=\dfrac{\gamma}{(1-\tilde{b})}$, $p_{1}^{(2)}=\dfrac{\gamma(\gamma-1+2\tilde{b})}{2(1-\tilde{b})^2}$, ${\mbox{\scriptsize{U}}_{1}^{(1)}}=\kappa_0\cos\beta$, ${\mbox{\scriptsize{V}}_{1}^{(1)}}={-\kappa_0\sin\beta}$,  ${\mbox{\scriptsize{U}}_{1}^{(2)}}=\dfrac{(\gamma-3+4\tilde{b})\kappa_0\cos\beta}{4(1-\tilde{b})}$, ${\mbox{\scriptsize{V}}_{1}^{(2)}}=\dfrac{(3-\gamma-4\tilde{b})\kappa_0\sin\beta}{4(1-\tilde{b})}$, $a_{1}^{(1)}=\dfrac{\kappa_0(\gamma-1+2\tilde{b})}{2(1-\tilde{b})}$, $a_{1}^{(2)}=\dfrac{\kappa_0((\gamma-1)(\gamma-3+8\tilde{b})+8\tilde{b}^2)}{8(1-\tilde{b})^2}$, ${\kappa_0}={(1-\tilde{b})^{-(\gamma+1)/2}}$, $c_0=a_0/{\kappa_0}$~ and ~$0<\beta<{\pi}$.\\
Similarly, for the asymptotic expansions of the state-2 variables, using (\ref{equ7*}) and $(\ref{equ42})_{1,2,3,4}$ into (\ref{equ11}), the R-H conditions (\ref{equ11}) on the reflected shock imply the following asymptotic forms:
\begin{align}\label{equ44}
\begin{split}
&\frac{\rho_2}{\rho_0}=1+{\rho_{2}^{(1)}}\epsilon+O(\epsilon^2),~~~~~~~~~~~~~~~~~~\frac{p_2}{p_0}=1+{p_{2}^{(1)}}\epsilon+p_{2}^{(2)}\epsilon^2+O(\epsilon^3),\\&
\frac{\mbox{\scriptsize{U}}_2}{c_0}={{\mbox{\scriptsize{U}}}_{2}^{(1)}}{\epsilon}+{{\mbox{\scriptsize{U}}}_{2}^{(2)}}{\epsilon}^2+O(\epsilon^3),~~~~~~~~~~~\frac{\mbox{\scriptsize{V}}_2}{c_0}={{\mbox{\scriptsize{V}}}_{2}^{(1)}}{\epsilon}+{{\mbox{\scriptsize{V}}}_{2}^{(2)}}{\epsilon}^2+O(\epsilon^3),
\end{split}
\end{align}
where perturbed quantities are as follows:\\
${\rho_{2}^{(1)}}=2$, ${p_{2}^{(1)}}=\dfrac{2\gamma}{(1-\tilde{b})}$, ${{\mbox{\scriptsize{U}}}_{2}^{(1)}}=0$, ${{\mbox{\scriptsize{V}}}_{2}^{(1)}}=0$, ${p_{2}^{(2)}}=\frac{\gamma}{{(1-\tilde{b})^2}}(2\gamma-1+3\tilde{b})$, ${{\mbox{\scriptsize{U}}}_{2}^{(2)}}=\dfrac{\kappa_0}{2(1-\tilde{b})}(-\gamma+1-2\tilde{b})\cos{\beta}$, ${{\mbox{\scriptsize{V}}}_{2}^{(2)}}=-\dfrac{\kappa_0}{2(1-\tilde{b})}(-\gamma+1-2\tilde{b})\sin{\beta}$,~ and ~${\pi}<\beta<{3\pi}/2$.
It follows from (\ref{equ6}), $(\ref{equ42})_1$ and $(\ref{equ44})_1$ that the solution, to the first order approximation, is piecewise constant in the exterior of the region $\tilde{\Omega}$, i.e.,
\begin{equation}\label{equ50}
{\rho_i^{(1)}({\zeta},\beta)} = \left\{
  \begin{array}{l l l}
	  {\rho}^{(1)}_0=0, & \quad (\zeta, \beta)\in{\Omega_0},\\ 
    {\rho}^{(1)}_1=1, & \quad (\zeta, \beta)\in{\Omega_1},\\
    {\rho}^{(1)}_2=2, & \quad (\zeta, \beta)\in{\Omega_2}.
\end{array} \right.
\end{equation}
%%%%%%%%%%%%%%%%%%%%%%%%%%%%%%%%%%%%%%%%%%%%%%%%%%%%%%%%%%%%%%%%%%%%%%%%%%%%%%%%%%%%%%%%%%%%%%%%%%%%%%%%%%%%%%%%%%%%%%%%%
\section{First order approximation in the diffracted region $\widetilde{\Omega}$}
In order to find the first order linear approximation to the problem, we look for asymptotic expansions of the form
\begin{align}\label{equ47}
\begin{split}
&{\rho}/{\rho_0}=1+{\epsilon}{\tilde{\rho}^{(1)}}+{{\epsilon}^2}{\tilde{\rho}^{(2)}}+O(\epsilon^3),\\&
{\mbox{\scriptsize{U}}}/{c_0}={\epsilon}{\kappa_0}\tilde{{\mbox{\scriptsize{U}}}}^{(1)}+{{\epsilon}^2}{\kappa_0}\tilde{{\mbox{\scriptsize{U}}}}^{(2)}+O(\epsilon^3),\\&
{\mbox{\scriptsize{V}}}/{c_0}={\epsilon}{\kappa_0}\tilde{{\mbox{\scriptsize{V}}}}^{(1)}+{{\epsilon}^2}{\kappa_0}\tilde{{\mbox{\scriptsize{V}}}}^{(2)}+O(\epsilon^3),\\&
({{\mbox{\scriptsize{S}}}-{\mbox{\scriptsize{S}}_0}})/{c_v}={\epsilon}\tilde{{\mbox{\scriptsize{S}}}}^{(1)}+{{\epsilon}^2}\tilde{{\mbox{\scriptsize{S}}}}^{(2)}+O(\epsilon^3).
\end{split}
\end{align}
 For convenience, introducing the non-dimensional variable $\xi=\zeta/{c_0}$ and inserting the asymptotic expansions (\ref{equ47}) into (\ref{equ8}), one gets the following system of equations for the first order perturbation variables  
\begin{align}\label{equ48*}
\begin{split}
&-{\xi^2}{\tilde{\rho}_{\xi}^{(1)}}+\kappa_0{\xi}{\tilde{{\mbox{\scriptsize{U}}}}_{\xi}^{(1)}}+\kappa_0(\tilde{{\mbox{\scriptsize{U}}}}^{(1)}+{\tilde{{\mbox{\scriptsize{V}}}}_{\beta}^{(1)}})=0,\\&
\kappa_0{\tilde{\rho}_{\xi}^{(1)}}-{\xi}{\tilde{\mbox{\scriptsize{U}}}_{\xi}^{(1)}}+(\kappa_0{(1-\tilde{b})}/{\gamma}){\tilde{\mbox{\scriptsize{S}}}_{\xi}^{(1)}}=0,\\&
-{\xi^2}{\tilde{\mbox{\scriptsize{V}}}_{\xi}^{(1)}}+\kappa_0{\tilde{\rho}_{\beta}^{(1)}}+(\kappa_0{(1-\tilde{b})}/{\gamma}){\tilde{\mbox{\scriptsize{S}}}_{\beta}^{(1)}}=0,\\&
{\tilde{\mbox{\scriptsize{S}}}_{\xi}^{(1)}}=0.
\end{split}
\end{align}
Eliminating $\tilde{{\mbox{\scriptsize{U}}}}^{(1)}$, $\tilde{{\mbox{\scriptsize{V}}}}^{(1)}$ and $\tilde{{\mbox{\scriptsize{S}}}}^{(1)}$from equations(\ref{equ48*}), one gets the following PDE in the unknown variable $\tilde{\rho}^{(1)}$
\begin{equation}\label{equ49}
{\xi^2}\left(\left({1-({\xi}/{\kappa_0})^2}\right){\tilde{\rho}_{\xi}^{(1)}}\right)_{\xi}+{\tilde{\rho}_{\beta\beta}^{(1)}}+{\xi}{\tilde{\rho}_{\xi}^{(1)}}=0.
\end{equation}
The change of radial variable $\xi$ to $r$, with $r=\frac{\xi/{\kappa_0}}{1+\sqrt{1-(\xi/{\kappa_0})^2}}$, transforms the PDE (\ref{equ49}) into the following Laplacian equation
\begin{equation}\label{equ50*}
r(r{\tilde{\rho}_{r}^{(1)}})_{r}+{\tilde{\rho}_{\beta{\beta}}^{(1)}}=0,
\end{equation}
the solution of which, following Keller and Blank (\cite{keller1951} with $\phi=\pi/4=\psi$ and $\lambda=2/3$), satisfying the boundary conditions (\ref{equ50}) and (\ref{equ7})can be written as 
\begin{equation}\label{equ51}
{\tilde{\rho}^{(1)}}=1+{\frac{1}{\pi}}\arctan\left\{{\frac{(1-r^{4/3})\sqrt{3}}{1+r^{4/3}+4r^{2/3}\cos(2\beta/3)}}\right\};~~~r\leq1,~~0\leq \beta \leq {3\pi}/2.
\end{equation}
It may be noticed that the solution given by (\ref{equ51}) is not uniformly valid throughout the domain $\tilde{\Omega}$, specially at the boundary $r=1$ or $\xi=\kappa_0$, where its derivatives blow up. Indeed, the normal derivative of linearized solution (\ref{equ51}) is unbounded in region $II$, while in region $I$, where the diffracted wave $S_D$ merges into reflected wave $S_R$ tangentially, both the normal and tangential derivatives of the linearized solution are unbounded. This means that it is essential to account for the nonlinear effects at those points where the singularity occurs. From (\ref{equ8}), it may be noticed that the system becomes degenerate at $\zeta={(\mbox{\scriptsize{U}}+a)}_{0}$. Therefore, we need a different asymptotic expansion at $\zeta={(\mbox{\scriptsize{U}}+a)}_{0}$ where the nonlinear effects are significant; this we discuss in the following section.  
The asymptotic behavior of (\ref{equ51}) near $\xi=\kappa_0$ is given by
\begin{equation}\label{equ52}
\tilde{{\rho}}^{(1)}=\rho_i^{(1)}({\zeta},\beta)+\left({\frac{2}{3}}\right)^{1/2}{\frac{2}{\pi(1+2\cos(2\beta/3))}}{\sqrt{1-\frac{\xi}{\kappa_0}}}+{O\left(1-\frac{\xi}{\kappa_0}\right)},
\end{equation}
which terminates at $Q(\kappa_0,\pi)$; therefore, we need a different asymptotic expansion in the neighborhood of $Q(\kappa_0, \pi)$ which we discuss in section 6.\\
One can construct higher order terms (say, ${\tilde{\rho}^{(2)}}$) in the expansion $(\ref{equ47})_1$. In view of the fact that ${\tilde{\rho}^{(2)}}=O(1-{\xi}/{\kappa_0})^{-1/2}$, the consecutive terms in this expansion would be progressively more singular at $r=1$; this renders the expansion $(\ref{equ47})_1$ to be non-uniform in regions $I$, $II$, and $IV$. 
Therefore, in view of (\ref{equ52}) and $(\ref{equ47})_1$, the first order approximation of the solution near $r=1$ for $\beta\neq \pi$ can be written as
\begin{equation}\label{equ53}
{\rho}/{\rho_0}=1+{\epsilon}\rho_i^{(1)}({\zeta},\beta)+\left(\frac{2}{3}\right)^{1/2}\frac{2\epsilon}{\pi(1+2\cos(2\beta/3))}{\sqrt{1-\frac{\xi}{\kappa_0}}}+O\left(\frac{\epsilon^2}{\sqrt{1-{\xi}/{\kappa_0}}}\right).
\end{equation}
From equation (\ref{equ53}), it may be observed that the first order solution given by (\ref{equ53}), is valid only if ${\epsilon^2}/{\sqrt{1-{\xi}/{\kappa_0}}}<<\epsilon$ i.e., $\epsilon^{2}<<1-{\xi}/{\kappa_0}<<1$ while near the boundary where the nonlinear effects are significant, it has a singularity of type $\sqrt{1-{\xi}/{\kappa_0}}$.
%%%%%%%%%%%%%%%%%%%%%%%%%%%%%%%%%%%%%%%%%%%%%%%%%%%%%%%%%%%%%%%%%%%%%%%%%%%%%%%%%%%%%%%%%%%%%%%%%%%%%%%%%%%%%%%%%%%%%%%%%%%%%%%%%%%%%%%%%%%%%%%%%%%%%
\section{ Nonlinear approximation}
It may be noticed that at $W_0$ $\zeta(W_0)={(\mbox{\scriptsize{U}} +a)}_{0}=a_0$, the system (\ref{equ8}) becomes degenerate  in the sense that $det(A-\zeta I) =0$ at ${W=W_0}$ (constant), and so the solution at $O(\epsilon)$, namely, $W_1$ might develop a singularity at this point as is evident from (\ref{equ53}). Thus, in order to account for the nonlinear effects near this point, where the singularity can occur, we need to construct a new expansion where $\zeta$ is close to $a_{0}$. When $\beta\neq \pi$, the radial variable $\zeta$, close to an eigenvalue $a_{0}$ of (\ref{equ8}), can be rescaled as $\zeta = \zeta(W_0)+\zeta'(W_0)(\mbox{\scriptsize{U}} +a-\zeta(W_0))+{\epsilon^2}\tau = c_0(\kappa_0 +\kappa_0 a_i^{(1)}\epsilon +\kappa_0 {\mbox{\scriptsize{U}}}_i^{(1)}\epsilon) + {\epsilon^2}\tau$, where $i=1$ and $2$ refer to the states ahead of the diffracted wavefront $PQ$ and behind the sonic arc $QR$, respectively. We, therefore, look for an asymptotic expansion valid in regions $II$, $III$, and $IV$ of the form:
\begin{align}\label{equ62}
\begin{split}
&\frac{\rho}{\rho_0}=1+\epsilon{\rho_{i}^{(1)}}+{\epsilon^2}{\hat{\rho}(\tau,\beta)}+{O(\epsilon^3)},\\&
\frac{\mbox{\scriptsize{U}}}{c_0}=\epsilon\kappa_0{\mbox{\scriptsize{U}}_{i}^{(1)}(\beta)}+{\epsilon^2}\kappa_0{\hat{\mbox{\scriptsize{U}}}(\tau,\beta)}+{O(\epsilon^3)},\\&
\frac{\mbox{\scriptsize{V}}}{c_0}=\epsilon\kappa_0{\mbox{\scriptsize{V}}_{i}^{(1)}(\beta)}+{\epsilon^2}\kappa_0{\hat{\mbox{\scriptsize{V}}}(\tau,\beta)}+{\epsilon^3}\kappa_0{\hat{\hat{\mbox{\scriptsize{V}}}}(\tau,\beta)}+{O(\epsilon^4)},\\&
\frac{a}{c_0}=\kappa_0+\kappa_0{a_{i}^{(1)}}\epsilon+\kappa_0{\hat{a}\epsilon^2}+O(\epsilon^3),\\&
\frac{\mbox{\scriptsize{S}}-{\mbox{\scriptsize{S}}}_0}{c_0}=\epsilon\kappa_0{\mbox{\scriptsize{S}}_{i}^{(1)}(\beta)}+{\epsilon^2}\kappa_0{\hat{\mbox{\scriptsize{S}}}(\tau,\beta)}+{O(\epsilon^3)},
\end{split}
\end{align}
where $\rho_{i}^{(1)}$, $\mbox{\scriptsize{U}}_{i}^{(1)}$, $\mbox{\scriptsize{V}}_{i}^{(1)}$, $a_{i}^{(1)}$ and $\mbox{\scriptsize{S}}_{i}^{(1)}$ are the coefficients of $\epsilon$ in (\ref{equ42}) or (\ref{equ44}) referring to states-1 or -2, respectively.
Equations (\ref{equ8}), in view of (\ref{equ62}), yield at $O(1)$ the following relations
\begin{equation}\label{equ63}
{\hat{\mbox{\scriptsize{S}}}}_{\tau}=0,~~\hat{\mbox{\scriptsize{U}}}-\mbox{\scriptsize{U}}_{i}^{(2)}(\beta)=({\hat{\rho}-\rho_{i}^{(2)}}),~~\hat{\mbox{\scriptsize{V}}}=\mbox{\scriptsize{V}}_{i}^{(2)}(\beta),
\end{equation}
where $\rho_{i}^{(2)}$, $\mbox{\scriptsize{U}}_{i}^{(2)}$, $\mbox{\scriptsize{V}}_{i}^{(2)}$ are the coefficients of $\epsilon^2$ in (\ref{equ42}) or (\ref{equ44}) referring to states-1 and -2, respectively.\\
Similarly to the orders $O(\epsilon)$ and $O(\epsilon^2)$, we have:
\begin{equation}\label{equ64}
O(\epsilon):~~~{a}_{i}^{(1)}={\rho}_{i}^{(1)},~~-a_{i}^{(1)}{\hat{\rho}}_{\tau}+{\rho_{i}^{(1)}}\hat{\mbox{\scriptsize{U}}}_{\tau}-{\mbox{\scriptsize{U}}_{i\beta}^{(1)}}\hat{\mbox{\scriptsize{V}}}_{\tau}=0,~~-\hat{\hat{\mbox{\scriptsize{V}}}}_{\tau}-\mbox{\scriptsize{U}}_{i\beta}^{(1)}{\hat{\rho}}_{\tau}=0,
\end{equation}
\begin{equation}\label{equ65}
O(\epsilon^2):~~~({\kappa_0}\hat{\mbox{\scriptsize{U}}}-{\tau})(\hat{\rho}+\hat{\mbox{\scriptsize{U}}})_{\tau}-\kappa_0{{\mbox{\scriptsize{V}}}_{i}^{(1)}\mbox{\scriptsize{U}}_{i\beta}^{(1)}}(\hat{\rho}+\hat{\mbox{\scriptsize{U}}})_{\tau}+{\hat{\mbox{\scriptsize{V}}}_{\beta}}-\kappa_0{\mbox{\scriptsize{U}}_{i\beta}^{(1)}\hat{\hat{\mbox{\scriptsize{V}}}}_{\tau}}+{\hat{\mbox{\scriptsize{U}}}}+2\hat{a}{\kappa_0}{\hat{\rho}}_{\tau}-{\kappa_0}\hat{\rho}{\hat{\rho}}_{\tau}+{\kappa_0}\hat{\rho}{\hat{\mbox{\scriptsize{U}}}}_{\tau}=0,
\end{equation}
where equation (\ref{equ65}) has been written on using the $O(\epsilon^2)$ equations obtained from $(\ref{equ8})_{1,2}$. 
The jump relations (\ref{equ42}) and (\ref{equ44}) yield the following relations
\begin{align}\label{equ66}
\begin{split}
&\mbox{\scriptsize{U}}_{i\beta}^{(1)}-\mbox{\scriptsize{V}}_{i}^{(1)}=\mbox{\scriptsize{U}}_{i}^{(1)}+\mbox{\scriptsize{V}}_{i\beta}^{(1)}=0,\\&
\hat{a}=\left\{\frac{(\gamma-1+2\tilde{b})}{2(1-\tilde{b})}\hat{\rho}+{\frac{(\gamma-1)(\gamma-3+{\gamma}\tilde{b}+5\tilde{b})}{8(1-\tilde{b})}}{\rho_{i}^{(1)}}^{2}+\frac{(\gamma+1)(\gamma+3)\tilde{b}^{2}{\rho_{i}^{(1)}}^{2}}{8(1-\tilde{b})^2}\right\}.
\end{split}
\end{align}
Now using (\ref{equ63}), (\ref{equ64}), and (\ref{equ66}) in (\ref{equ65}) we get
\begin{equation}\label{equ67}
\left\{{\kappa_0}\left(\frac{\gamma+1}{1-\tilde{b}}\right)({\hat{\rho}-\rho_{i}^{(2)}})-2({\tau}-K)\right\}\hat{\rho}_{\tau}+\hat{\rho}-\rho_{i}^{(2)}=0,
\end{equation}
where
\begin{align}
\begin{split}
%&\tilde{\rho}=\hat{\rho}-\rho^{(2)}\\&
 &K={\kappa_0}\mbox{\scriptsize{U}}_{i}^{(2)}+\frac{{\kappa_0}{\rho_{i}^{(2)}}}{2}\left(\frac{\gamma-1+2\tilde{b}}{1-\tilde{b}}\right)-\frac{\kappa_0}{2}\mbox{\scriptsize{V}}_{i}^{(1)}\mbox{\scriptsize{U}}_{i\beta}^{(1)}\\&
+{\kappa_0}\left\{\frac{(\gamma-1)(\gamma-3+{\gamma}\tilde{b}+5\tilde{b})}{8(1-\tilde{b})}{\rho_{i}^{(1)}}^2+\frac{(\gamma+1)(\gamma+3)\tilde{b}^2{\rho_{i}^{(1)}}^2}{8(1-\tilde{b})^2}\right\}.
\end{split}
\end{align}
Solution for (\ref{equ67}) is given by
\begin{equation}\label{equ68}
{C}(\hat{\rho}-\rho_{i}^{(2)})^{2}+{\kappa_0}\left(\frac{\gamma+1}{1-\tilde{b}}\right)(\hat{\rho}-\rho_{i}^{(2)})-({\tau}-K)=0,
\end{equation}
where $C=C(\beta)$ is the integration constant. Thus, depending on the sign of integration constant $C(\beta)$ of (\ref{equ68}), there are two possible solutions; the first possible solution, for $C>0$, is a parabola in $(\tau, \hat{\rho})$-plane with vertex in the second quadrant and branches in $+\tau$ direction. For a uniformly valid solution, boundary conditions must be satisfied, i.e., $\hat{\rho}=\rho_{i}^{(2)}$ as ${\tau}\rightarrow\infty$; but in this case the solution $\hat{\rho}$ takes either lower or upper branch of parabola as  ${\tau}\rightarrow\infty$. This shows that $C$ cannot be positive; indeed, it must be negative for the desired unique solution. Thus, there exists a value of $\tau$, say $\tau_s$ such that for $\tau < \tau_s$, there may be a shock or an expansion wave, whereas for $\tau>\tau_s$, $\hat{\rho}=\rho_{i}^{(2)}$. In case of a shock, R-H conditions along with the entropy condition must be satisfied; equation (\ref{equ67}) can be expressed in the following conservation form
\begin{equation}\label{equ69}
\left\{{\kappa_0}\left(\frac{\gamma+1}{1-\tilde{b}}\right)\frac{(\hat{\rho}-\rho_{i}^{(2)})^2}{2}-2({\tau}-K)({\hat{\rho}}-\rho_{i}^{(2)})\right\}_{{\tau}}+3(\hat{\rho}-\rho_{i}^{(2)})=0,
\end{equation}
which yields the R-H condition
\begin{equation}\label{equ70}
\left[{\kappa_0}\left(\frac{\gamma+1}{1-\tilde{b}}\right)\frac{(\hat{\rho}-\rho_{i}^{(2)})^2}{2}-2({\tau}-K){(\hat{\rho}-\rho_{i}^{(2)})}\right]_{{\tau}={\tau}_s}=0.
\end{equation}
The entropy condition takes the form  $\hat{\rho}{{(\tau_s}^{+})}<\hat{\rho}{({\tau_s}^{-})}$; as $\hat{\rho}{{(\tau_s}^{+})}=\rho_{i}^{(2)}$  is always positive, we have $\hat{\rho}{({\tau_s}^{-})}>0$. Moreover, it follows from the jump condition (\ref{equ70}) and the entropy condition that
\begin{equation}\label{equ71}
\hat{\rho}({\tau_s}^{-})=\frac{4}{\kappa_0}\left(\frac{1-\tilde{b}}{\gamma+1}\right)({\tau_s}-K)+\rho_{i}^{(2)}>0,
\end{equation}
which, in view of (\ref{equ68}), implies that for shocks
\begin{equation}\label{equ72}
{\tau_s}=\frac{-3}{C(\beta)}\left\{\frac{\kappa_0}{4}\left(\frac{\gamma+1}{1-\tilde{b}}\right)\right\}^{2}+K.
\end{equation}
Thus, one can uniquely determine the solution $\hat{\rho}$ from (\ref{equ68}) if $C(\beta)$ is known. It follows from (\ref{equ68})  that for a shock, the function $\hat{\rho}$ will take the upper branch of the parabola, however for a continuous solution, $\hat{\rho}$ will take the lower branch (see Figure 2).
\begin{figure}
\centering
\includegraphics[width=4.5in]{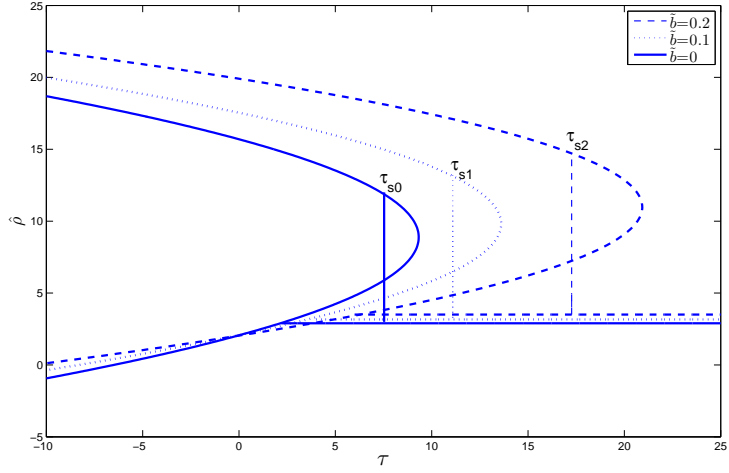}
\caption{Solution profile $\hat{\rho}({\tau})$ and the shock location influenced by the parameter $\tilde{b}$ for different values of $\tilde{b}$.}
\end{figure}
By matching the non-linear approximation $(\ref{equ62})_1$ with the linear approximation $(\ref{equ47})_1$, we obtain the value of $\hat{\rho}$ and the appropriate branch of the parabola. It follows from (\ref{equ68}) that $\hat{\rho}$ expressed as
\begin{equation}\label{equ72*}
\hat{\rho}\approx\pm\sqrt{\dfrac{\tau}{C(\beta)}},~~~~~~ \text{as}~~~ \tau\longrightarrow-\infty
\end{equation}
In view of (\ref{equ72*}) and the expansion for $\tau$, the non-linear approximation  $(\ref{equ62})_1$ yields
\begin{align}\label{equ73}
\begin{split}
\frac{\rho}{\rho_0}&=1+\epsilon{\rho_{i}^{(1)}}+{{\epsilon}^2}\left\{\pm\sqrt{\frac{\kappa_0}{C(\beta)}\left(\left(\dfrac{\zeta}{{\kappa_0} c_0}-1\right)\frac{1}{\epsilon^2}+O\left(\dfrac{1}{\epsilon}\right)\right)}\right\}\\&
 =1+\epsilon{\rho_{i}^{(1)}}\pm{\epsilon}\sqrt{\frac{\kappa_0}{C(\beta)}\left(\dfrac{\zeta}{{\kappa_0} c_0}-1\right)}+O({\epsilon}^2).
\end{split}
\end{align}
From $(\ref{equ47})_1$ and  $(\ref{equ52})$, one finds that
\begin{align}\label{equ74}
\begin{split}
\frac{\rho}{\rho_0}=&1+{\epsilon}\rho_i^{(1)}({\zeta},\beta)+\left(\frac{2}{3}\right)^{1/2}\frac{2\epsilon}{\pi(1+2\cos(2\beta/3))}{\sqrt{1-\frac{\zeta}{\kappa_0 c_0}}}\\&+\epsilon{O\left(1-\frac{\zeta}{\kappa_0 c_0}\right)}+\epsilon^{2}O\left(\frac{1}{\sqrt{1-\frac{\zeta}{\kappa_0 c_0}}}\right)+O(\epsilon^3).
\end{split}
\end{align}
Comparing the coefficient of $O(\epsilon)$ from (\ref{equ73}) and (\ref{equ74}), we get
\begin{equation}\label{equ75*}
\pm\sqrt{\dfrac{\kappa_0}{-C(\beta)}}=\left(\frac{2}{3}\right)^{1/2}\frac{2}{\pi(1+2\cos(2\beta/3))},
\end{equation}
which is possible only if
\begin{equation*}
\epsilon^{2}O\left(\frac{1}{\sqrt{1-\dfrac{\zeta}{\kappa_0 c_0}}}\right)<<\epsilon{\sqrt{1-\frac{\zeta}{\kappa_0 c_0}}}
\end{equation*}
i.e.,
\begin{equation*}
\epsilon<<{1-\frac{\zeta}{\kappa_0 c_0}}<<1.
\end{equation*}
Therefore, from (\ref{equ75*})
\begin{equation}\label{equ77}
\left(\frac{2}{3}\right)^{1/2}\frac{2}{\pi(1+2\cos(2\beta/3))} = \left\{
  \begin{array}{l l}
    \sqrt{\dfrac{\kappa_0}{-C(\beta)}}, & {\beta}<{\pi},\\
    -\sqrt{\dfrac{\kappa_0}{{-C(\beta)}}}, & \beta>\pi,
\end{array} \right.
\end{equation}
since
\begin{equation*}
{(1+2\cos(2\beta/3))} = \left\{
  \begin{array}{l l}
    {>0}, & {\beta}<{\pi},\\
    {<0}, & \beta>\pi,
\end{array} \right.
\end{equation*}
On squaring(\ref{equ75*}), we obtain the value of the unknown $C(\beta)$ as
\begin{equation}\label{equ76*}
C(\beta)=\frac{-3\kappa_0{\pi}^2(1+2\cos{2\beta/3})^2}{8}.
\end{equation}
In view of (\ref{equ76*}), equation (\ref{equ72}) gives the expression for the shock location for $\beta\neq \pi$ 
\begin{equation}\label{equ77*}
\tau_s=\frac{(\gamma+1)^2}{2{\pi}^2(1-\tilde{b})^{{\gamma+5}/2}(1+2\cos(2\beta/3))^2}+K.
\end{equation}
 Equation (\ref{equ77*}) shows that the location of the diffracted shock, in self-similar polar coordinates, which is same as the velocity of diffracted shock in polar coordinates-$(r, t)$, is significantly affected by the van der Waals excluded volume $\tilde{b}$. Indeed, the diffracted shock in a real gas $(0<\tilde{b}<1)$ moves faster as compared to the ideal gas case $(\tilde{b}=0)$.\\
From (\ref{equ68}) $\hat{\rho}$ is given by
\begin{equation}\label{equ79}
{[\rho]}\equiv{\hat{\rho}-\rho_{i}^{(2)}} = \left\{
  \begin{array}{l l}
    {\dfrac{-(\gamma+1){\kappa_0}+\sqrt{(\gamma+1)^2 {\kappa_0}^2+4C(\tau-K)(1-\tilde{b})^2}}{2C(\beta)(1-\tilde{b})}}, & {\beta}<{\pi},~~~ \text{(shock)}\\
    {\dfrac{-(\gamma+1){\kappa_0}-\sqrt{(\gamma+1)^2 {\kappa_0}^2+4C(\tau-K)(1-\tilde{b})^2}}{2C(\beta)(1-\tilde{b})}}, & \beta>\pi. ~~~\text{(expansion)}\\
\end{array} \right.
\end{equation} 
Here, we take positive root for the diffracted shock $PQ$ and negative for the expansion wave $QR$ since across the shock density increases while across expansion wave it decreases.
Now, using (\ref{equ77*}) and (\ref{equ76*}) in $(\ref{equ79})_1$ and matching with the boundary condition $(\ref{equ42})_1$, the shock strength across the diffracted shock is given as
\begin{equation}\label{equ77**}
[\rho]=\frac{2(\gamma+1)}{3\pi^2(1-\tilde{b})(1+2\cos(2\beta/3))^2}.
\end{equation}
 Equation (\ref{equ77**}) shows that the diffracted shock $PQ$ is stronger in a non-ideal gas $(0<\tilde{b}<1)$ as compared to the ideal gas $(\tilde{b}=0)$ case. Similarly, in view of $(\ref{equ76*})$ and the boundary condition $(\ref{equ44})_1$,  jump in the density gradient across the expansion wave $QR$ is obtained from $(\ref{equ79})_2$ along the radial direction to be
\begin{equation}\label{equ79*}
[\rho_{\tau}]=\frac{2(1-\tilde{b})^{\gamma+3/2}}{\gamma+1},
\end{equation}
which shows that an increase in $\tilde{b}$ causes the density gradient to decrease, implying thereby that the rarefaction wave becomes weaker and decays slowly as compared to the corresponding ideal gas ($\tilde{b}=0$) case.
%%%%%%%%%%%%%%%%%%%%%%%%%%%%%%%%%%%%%%%%%%%%%%%%%%%%%%%%%%%%%%%%%%%%%%%%%%%%%%%%%%%%%%%%%%%%%%%%%%%%%%%%%%%%%5
\section{Asymptotic approximation in the neighborhood of $Q(a_0,\pi)$ }
It may be noticed that the linearized solution given by (\ref{equ52}), is invalid in the neighborhood of $Q$. The purpose of this section is to derive equations which are valid in region $I$ (see Figure 1). In order to find a complete asymptotic description of the shock diffraction problem we construct asymptotic expansions valid in region $I$ by stretching the variables $\xi$ and $\beta$ and match leading order solutions in different regions, shown in Figure 1; to this end we introduce new variables $r^{'}=(\xi-\kappa_0)/{\epsilon}$ and $\beta^{'}=(\beta-\pi)/{\epsilon^{\Delta}}$, where $\epsilon^{\Delta}$ is the gauge function with $\Delta>0$ to be determined. In terms of these new variables, the dominant part of (\ref{equ49}), after simplification, results into
\begin{equation}\label{equ52*}
2{\kappa_0}r^{'}\tilde{\rho}_{r^{'}r^{'}}^{(1)}+{\kappa_0}\tilde{\rho}_{r^{'}}^{(1)}-\epsilon^{1-2\Delta}\tilde{\rho}_{\beta^{'}\beta^{'}}^{(1)}=0.
\end{equation}
At this point, based on the principle that the leading order equation should be kept as rich as possible so that the solution contains the maximum possible information, the only choice for $\Delta$, which gives a non-degenerate reduced problem and allows all the terms in  (\ref{equ52*}) to be retained is $\Delta=1/2$. In terms of these new variables (\ref{equ51}) yields
\begin{equation}\label{equ53*}
\tilde{{\rho}}^{(1)}=2-{\frac{1}{\pi}}\tan^{-1}{\frac{\sqrt{-2r'/\kappa_0}}{\beta'}}+O(\epsilon^{1/2}),
\end{equation}
where $r'<0$. \\
Accordingly, in region $I$, we look for the asymptotic expansions of the form:
\begin{align}\label{equ80}
\begin{split}
&{\rho}/{\rho_0}=1+{\epsilon}{\bar{\rho}(r',{\beta'})}+{\epsilon}^{3/2}{\bar{\bar{\rho}}(r',{\beta'})}+{O(\epsilon^{2})},\\&
{\mbox{\scriptsize{U}}}/{c_0}={\epsilon}{\kappa_0}{\bar{\mbox{\scriptsize{U}}}(r',{\beta'})}+{\epsilon}^{3/2}{\kappa_0}{\bar{\bar{\mbox{\scriptsize{U}}}}(r',{\beta'})}+{O(\epsilon^{2})},\\&
{\mbox{\scriptsize{V}}}/{c_0}={\epsilon}{\kappa_0}{\bar{\mbox{\scriptsize{V}}}(r',{\beta'})}+{\epsilon^{3/2}}{\kappa_0}{\bar{\bar{\mbox{\scriptsize{V}}}}(r',{\beta'})}+{O(\epsilon^{2})},\\&
{a}/{c_0}=\kappa_0+{\epsilon}\kappa_0\bar{a}+\epsilon^{3/2}\kappa_0{\bar{a}}+O(\epsilon^2),\\&
\frac{{\mbox{\scriptsize{S}}}-{\mbox{\scriptsize{S}}_0}}{c_v}={\epsilon}{\bar{\mbox{\scriptsize{S}}}(r',{\beta'})}+{\epsilon}^{3/2}{\bar{\bar{\mbox{\scriptsize{S}}}}(r',{\beta'})}+{O(\epsilon^{2})}.
\end{split}
\end{align}
When the expansions (\ref{equ80}) are inserted into the equations (\ref{equ8}), and collecting respectively, $O(1)$, $O(\epsilon^{1/2})$ and $O(\epsilon)$ terms, there results the following system of equations
\begin{equation}\label{equ81}
O(1):~~~~{\bar{\rho}_{r'}}=\bar{\mbox{\scriptsize{U}}}_{r'}\Rightarrow \bar{\mbox{\scriptsize{U}}}-\mbox{\scriptsize{U}}_{i}^{(1)}=\bar{\rho}-\rho_{i}^{(1)},~~\bar{\mbox{\scriptsize{V}}}_{r'}=0,~~\bar{\mbox{\scriptsize{S}}}_{r'}=0.
\end{equation}
\begin{equation}\label{equ82}
~~~~~~O(\epsilon^{1/2}):~~~~\bar{\mbox{\scriptsize{V}}}_{\beta^{'}}=0,~~~{\kappa_0}\bar{\bar{\mbox{\scriptsize{V}}}}_{r'}=\bar{\rho}_{\beta^{'}},~~~\bar{\bar{\mbox{\scriptsize{S}}}}_{r'}=0.
\end{equation}
\begin{equation}\label{equ83}
O(\epsilon):~~~~({\kappa_0}\bar{\mbox{\scriptsize{U}}}-r')\bar{\rho}_{r'}+{\kappa_0}\bar{\rho}\bar{\mbox{\scriptsize{U}}}_{r'}+\bar{\mbox{\scriptsize{U}}}+\bar{\bar{\mbox{\scriptsize{V}}}}_{\beta^{'}}=0,~~{\kappa_0}(2\bar{a}-\bar{\rho})\bar{\rho}_{r'}+({\kappa_0}\bar{\mbox{\scriptsize{U}}}-r')\bar{\mbox{\scriptsize{U}}}_{r'}=0.
\end{equation}
In view of $(\ref{equ81})_1$, combining equations $(\ref{equ83})_1$ and $(\ref{equ83})_2$, we get
%\begin{equation}\label{equ83} 
%{M}{X}+{N}\bar{R}=0, 
%\end{equation}
%where ${X}=\bar{\bar{W}}_{r'}/{U_{\theta'}}$.
\begin{equation}\label{equ84}
2({\kappa_0}(\bar{\rho}-\rho_{i}^{(1)}+\mbox{\scriptsize{U}}_{i}^{(1)})-r'+{\kappa_0}\bar{a})\bar{\rho}_{r'}+\bar{\bar{\mbox{\scriptsize{V}}}}_{\beta^{'}}+\bar{\rho}-\rho_{i}^{(1)}+\mbox{\scriptsize{U}}_{i}^{(1)}=0,
\end{equation}
where $\bar{a}$ is the coefficient of $\epsilon$ in  $(\ref{equ80})_4$ given by the following relation
\begin{equation}\label{equ85}
 \bar{a}=\frac{(\gamma-1+2\tilde{b})\bar{\rho}}{2(1-\tilde{b})}.
 \end{equation}
 Now, using (\ref{equ85}) in (\ref{equ84}), and writing the resulting equation and $(\ref{equ82})_2$ in divergence form, we get
 \begin{equation}\label{equ86}
 2\left\{\frac{\kappa_0(\gamma+1)\bar{\rho}}{2(1-\tilde{b})}-\left(r'-\kappa_0(\mbox{\scriptsize{U}}_{i}^{(1)}-\rho_{i}^{(1)})\right)\right\}\bar{\rho}_{r'}+\bar{\rho}-\rho_{i}^{(1)}+\mbox{\scriptsize{U}}_{i}^{(1)}+\bar{\bar{\mbox{\scriptsize{V}}}}_{\beta^{'}}=0,~~~~~{\kappa_0}\bar{\bar{\mbox{\scriptsize{V}}}}_{r'}=\bar{\rho}_{\beta^{'}}.
 \end{equation}
Let $\tilde{r}=S_R(\beta')$ be the location of the reflected shock in region $I$. Then the shock conditions for (\ref{equ86}) are
\begin{equation}\label{equ90}
 \kappa_0[\bar{\bar{\mbox{\scriptsize{V}}}}]+(d{S_{R}}/{d\beta^{'}})[\bar{\rho}]=0,
\end{equation}
\begin{equation}\label{equ91}
{\vartheta}[\bar{\rho}^2]-(d{S_{R}}/{d\beta^{'}})[\bar{\bar{\mbox{\scriptsize{V}}}}]-2 S_{R}[\bar{\rho}]=0,
\end{equation}
where $\vartheta=(\kappa_0/2)(\gamma+1)(1-\tilde{b})^{-1}$ and $\tilde{r}=r'-\kappa_0(\mbox{\scriptsize{U}}_{i}^{(1)}-\rho_{i}^{(1)})$.\\
Eliminating $[\bar{\bar{\mbox{\scriptsize{V}}}}]$ from (\ref{equ90}) and (\ref{equ91}), we get
\begin{equation}\label{equ91*}
{2\kappa_0}{\vartheta}<\bar{\rho}>+(d{S_{R}}/{d\beta^{'}})^{2}-2\kappa_0 S_{R}=0,
\end{equation}
where $<\bar{\rho}>$ denotes the average value of $\bar{\rho}$ on either side of the reflected shock.
The solution of the above equation, in view of the fact that the values of $\bar{\rho}$ ahead and behind of $S_R$ are 1 and 2, respectively, can be written as 
\begin{equation}\label{equ92}
S_R=(\kappa_0/2)(\beta'-\beta_0)^2+(3\kappa_0/4)(\gamma+1)/(1-\tilde{b}),
\end{equation}
where $\beta_0$ is an arbitrary constant.
In order to find a uniformly valid solution throughout the flow field we find matching conditions for (\ref{equ86}). First, $(\ref{equ80})_1$ must match with the solution in region $III$, given by $(\ref{equ50})_{2,3}$. This implies that a weak solution for the system  (\ref{equ86}) can be written as
\begin{equation}\label{equ93}
 {\bar{\rho}(\tilde{r}, \beta')}
  = \left\{
  \begin{array}{l l}
    1, & \tilde{r}>S_R,\\
    2, & \tilde{r}<S_R.
\end{array} \right.
\end{equation}
 Since the stretched variables in regions $\tilde{\Omega}$ and $I$ are related by $\zeta=c_0(\kappa_0+\epsilon r')$ and $\beta=\pi+\epsilon^{1/2}\beta'$, the matching condition for the leading order solutions in regions $\tilde{\Omega}$ and $I$ can be obtained in conformity with (\ref{equ42}), (\ref{equ44}), and $(\ref{equ47})_1$ as follows:\\
From (\ref{equ7*}), we get the following approximations for the reflected shock in $(r', \beta')$ plane 
\begin{equation}\label{equ88}
r'=\frac{\kappa_0{\beta'}^2}{2} ~~~~~~\text{as}~~\beta'\rightarrow{\infty},
\end{equation}
showing thereby that, in region $I$, the straight reflected shock becomes parabolic in the limit $\beta'\rightarrow{\infty}$; we notice that an increase in the van der Waals excluded volume $\tilde{b}$ causes an increase in its latus-rectum, indicating thereby that an increase in $\tilde{b}$ causes the real gas boundaries to become larger than the corresponding ideal gas case. In view of (\ref{equ42}), (\ref{equ44}), (\ref{equ53*}), and (\ref{equ88}), the boundary conditions for (\ref{equ86}) can be specified as
\begin{equation}\label{equ89}
 \lim_{{\beta'} \to \infty}{\bar{\rho}(\eta{\kappa_0}\frac{{\beta'}^2}{2}, \beta')}\approx \lim_{{\epsilon} \to 0}{\tilde{\rho}^{(1)}}
  = \left\{
  \begin{array}{l l l}
    1, & {\eta}>1,\\
    2, & 0<{\eta}<1,\\
		{2-\frac{1}{\pi}\tan^{-1}{\sqrt{-\eta}}}, & {\eta}<0.
\end{array} \right.
\end{equation}
In a similar manner the solution in region $I$ must match with the solution in region $II$. Let $\tilde{r}=S_D(\beta')$ be the equation of the diffracted shock; then in view of (\ref{equ86}), the location of the diffracted shock can be obtained in the following form
\begin{equation}\label{equ94}
S_D=(\kappa_0/2)(\beta'-\beta_0)^2+(\kappa_0/4)(\gamma+1)(3-1/\pi \tan^{-1}\sqrt{-\eta})/(1-\tilde{b}),
\end{equation}
and the diffracted wave solution of the system (\ref{equ86}) satisfying the boundary conditions $(\ref{equ89})_{1,3}$ can be written as
\begin{equation}\label{equ95}
 {\bar{\rho}(\tilde{r}, \beta')}
  = \left\{
  \begin{array}{l l}
    1, & \tilde{r}>S_D,\\
    {2-\frac{1}{\pi}\tan^{-1}{\sqrt{-\eta}}}, & \tilde{r}<S_D.
\end{array} \right.
\end{equation}
Now for smooth solution, (\ref{equ86}) becomes
\begin{equation}\label{equ87}
\kappa_0(\vartheta{\bar{\rho}}-\tilde{r}){\bar{\rho}}_{\tilde{r}\tilde{r}}+\kappa_{0}{\vartheta}\bar{\rho}_{\tilde{r}}^{2}-{\kappa_0}\bar{\rho}_{\tilde{r}}+\bar{\rho}_{\beta^{'}\beta^{'}}=0.
\end{equation}
 It may be remarked that the equation (\ref{equ87}) is a PDE in $\bar{\rho}$ implying the first approximation to the flow in the vicinity of the point $Q$.
It may be noticed that equation (\ref{equ87})  is of mixed type, namely, it is hyperbolic when $\vartheta \bar{\rho}<\tilde{r}$, and elliptic when $\vartheta \bar{\rho}>\tilde{r}$; however, when $\vartheta \bar{\rho}=\tilde{r}$, it corresponds to two sonic lines, $S_1:\tilde{r}= \vartheta$ and $S_2:\tilde{r}=2\vartheta$. Indeed, at $\tilde{r}=2\vartheta$ in region $I$, the reflected shock starts bending and merges asymptotically into the diffracted shock $S_D$; however the sonic line $\tilde{r}=\vartheta$ is asymptotic to the diffracted shock $S_D$  in the neighborhood of the point $Q'$ (see Figure $3$). \\
It may be observed that the stretching transformation $\tilde{r}\rightarrow {g^2}\tilde{r}$, $\beta'\rightarrow {g}\beta'$, ${\bar{\rho}}\rightarrow {g^2}\bar{\rho}$, for every parameter $g>0$, leaves the equation (\ref{equ87}) invariant and, therefore, it admits a similarity solution of the form $\bar{\rho}={\beta'}^2 f(\tilde{r}/{\beta'}^2)$ such that
\begin{equation}\label{equ96}
(4m^2+(2\kappa_0/{\beta'}^2)(\vartheta{\beta'}^2f-\tilde{r}))f''-(\kappa_0+2m)f'+2\kappa_0(f')^2+2f=0,
\end{equation}
with $m={\tilde{r}}/{\beta'}^2$; further, as the homogeneous equation (\ref{equ96}) admits a solution of the form $f(m)=\sqrt{m}$, an expansion wave solution of (\ref{equ87}) in the region $E$ between sonic lines $\tilde{r}=\vartheta$ and $\tilde{r}=2\vartheta$ and satisfying the boundary conditions $(\ref{equ89})_{2,3}$ can be written as
\begin{equation}\label{equ98}
{\bar{\rho}} = \left\{
  \begin{array}{l l l}
	  {2}, & m> 2\vartheta/{\beta'}^2.\\
    {{\beta'}^2 \sqrt{m}}, &\vartheta/{\beta'}^2 <m< 2\vartheta/{\beta'}^2,\\
    {2-\frac{1}{\pi}\tan^{-1}{\sqrt{-\eta}}}, &m< \vartheta/{\beta'}^2,
\end{array} \right.
\end{equation}
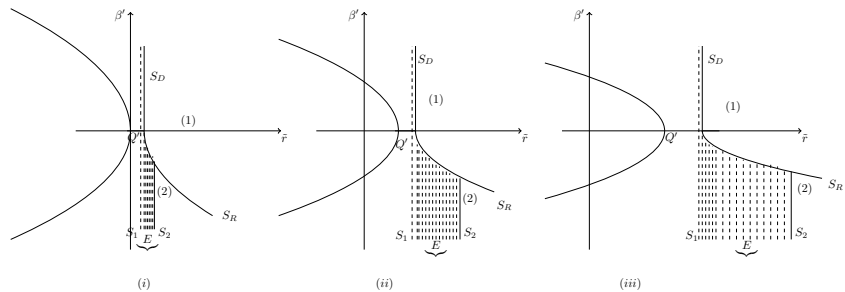
\begin{figure}[h]
\centering
\scalebox{.45}{
\begin{tikzpicture}
			\begin{scope}
			\draw [->][thick](-4,0)--(2,0);
			 \draw [->][thick](-2.4,-3.5)--(-2.4,3.5);
	  \draw [thick](-2,0)--(-2,2.5);
	  \draw [dashed](-2,-2.9)--(-2,0);
	  \draw [thick,dashed](-2.1,-2.9)--(-2.1,2.5);
	  
\draw [thick,-][rotate around={90:(-2.4,0)}] (-5.6,3.5) parabola bend (-2.4,0) (1.2,3.5);
\draw [-][thick](-1.7,-2.9)--(-1.7,-0.9);
       %\draw [rotate around={-90:(-1.5,-1)}][red,thick,-] (-1.5,-1) parabola  (0,1.5);
       \draw [rotate around={-90:(-2,0)}][thick,-] (-2,-0.5) parabola bend (-2,0)  (0.5,2);
       \draw [dashed](-1.75,-2.9)--(-1.75,-.75);
       \draw [dashed](-1.8,-2.9)--(-1.8,-.8);
       \draw [dashed](-1.85,-2.9)--(-1.85,-.7);
       \draw [dashed](-1.9,-2.9)--(-1.9,-.5);
       \draw [dashed](-1.95,-2.9)--(-1.95,-.4);
       %\draw [thick,-](3.3,4.6)--(3.8,4.6);
       \end{scope}
	\draw (2.1,-.2) node {$\tilde{r}$};	
	\draw (-2.7,3.4) node {$\beta'$};
	\draw (-2.3,-.2) node {$Q'$};
	%\draw (-1.4,-.8) node {$D'$};
	\draw (0.5,-2.5) node {$S_R$};
	\draw (-1.6,1.6) node {$S_D$};
	\draw (-2.35,-3) node {$S_1$};
  \draw (-1.4,-3) node {$S_2$};
  \draw (-.7,0.3) node {$(1)$};
  \draw (-1.4,-1.8) node {$(2)$};
  \draw (-2,-4.5) node {$(i)$};
  \draw (-1.9,-3.3) node {$\underbrace{E}$};
   \end{tikzpicture}\hspace{-.5cm}
   \begin{tikzpicture}
			\begin{scope}
			\draw [->][thick](-4,0)--(2,0);
	  \draw [->][thick](-2.6,-3.5)--(-2.6,3.5);
   \draw [thick,-][rotate around={90:(-1.6,0)}] (-4.1,3.5) parabola bend (-1.6,0) (1.1,3.5);
   \draw [rotate around={-90:(-1.1,0)}][thick,-] (-1.1,-0.6) parabola bend (-1.1,0)  (0.7,2.3);
       \draw [thick,-](-1.1,0)--(-1.1,2.5);
       \draw [thick,dashed](-1.2,-3.2)--(-1.2,2.5);
       \draw [thick,-](0.2,-3.2)--(0.2,-1.4);
       \draw [dashed](0.1,-3.2)--(0.1,-1.35);
      \draw [dashed](0,-3.2)--(0,-1.3);
      \draw [dashed](-0.1,-3.2)--(-0.1,-1.25);
      \draw [dashed](-0.2,-3.2)--(-0.2,-1.2);
      \draw [dashed](-0.3,-3.2)--(-0.3,-1.1);
      \draw [dashed](-0.4,-3.2)--(-0.4,-1.05);
      \draw [dashed](-0.5,-3.2)--(-0.5,-1);
      \draw [dashed](-0.6,-3.2)--(-0.6,-.9);
      \draw [dashed](-0.7,-3.2)--(-0.7,-.8);
      \draw [dashed](-0.8,-3.2)--(-0.8,-.7);
      \draw [dashed](-0.9,-3.2)--(-0.9,-.6);
      \draw [dashed](-1,-3.2)--(-1,-.5);
      \draw [dashed](-1.05,-3.2)--(-1.05,-.4);
      \end{scope}
      \draw (2.1,-.2) node {$\tilde{r}$};	
	\draw (-2.9,3.4) node {$\beta'$};
      \draw (-1.5,-.4) node {$Q'$};
      %\draw (.4,-1.1) node {$D'$};
      \draw (1.5,-2.2) node {$S_R$};
      \draw (-.8,2.1) node {$S_D$};
      \draw (-1.5,-3.1) node {$S_1$};
  		\draw (0.5,-3) node {$S_2$};
  		\draw (-.5,.9) node {$(1)$};
  		\draw (.5,-2) node {$(2)$};
  		\draw (-2,-4.5) node {$(ii)$};
  		\draw (-.5,-3.5) node {$\underbrace{E}$};
  		\end{tikzpicture}\hspace{.25cm}
  		\begin{tikzpicture}
			\begin{scope}
			\draw [->][thick](-3,0)--(4,0);
	  \draw [->][thick](-2.2,-3.5)--(-2.2,3.5);
  	\draw [thick,-][rotate around={90:(0,0)}] (-2,3.5) parabola bend (0,0) (2,3.5);	
  	\draw [rotate around={-90:(1.1,0)}][thick,-] (1.1,0.5) parabola bend (1.1,0) (2.5,3.5);
      \draw [thick,-](1.1,0)--(1.1,2.5);
      \draw [thick,dashed](1,-3.2)--(1,2.5);
  	 	 \draw [thick,-](3.7,-3.2)--(3.7,-1.2);
  	 	 \draw [dashed](3.5,-3.2)--(3.5,-1.1);
  	 	 \draw [dashed](3.3,-3.2)--(3.3,-1.05);
  	 	 \draw [dashed](3.1,-3.2)--(3.1,-1);
  	 	 \draw [dashed](2.9,-3.2)--(2.9,-1);
  	 	 \draw [dashed](2.7,-3.2)--(2.7,-.95);
  	 	 \draw [dashed](2.5,-3.2)--(2.5,-.9);
  	 	 \draw [dashed](2.3,-3.2)--(2.3,-.85);
  	 	 \draw [dashed](2.1,-3.2)--(2.1,-.8);
  	 	 \draw [dashed](1.9,-3.2)--(1.9,-.75);
  	 	 \draw [dashed](1.7,-3.2)--(1.7,-.6);
  	 	 \draw [dashed](1.5,-3.2)--(1.5,-.5);
  	 	 \draw [dashed](1.4,-3.2)--(1.4,-.45);
  	 	 \draw [dashed](1.3,-3.2)--(1.3,-.4);
  	 	 \draw [dashed](1.2,-3.2)--(1.2,-.3);
  	 	 \draw [dashed](1.1,-3.2)--(1.1,0);
  	 	 	\end{scope}
	\draw (4.1,-.2) node {$\tilde{r}$};	
	\draw (-2.5,3.4) node {$\beta'$};
	\draw (0.2,-.2) node {$Q'$};
	%\draw (3.8,-1) node {$D'$};
	\draw (5,-1.6) node {$S_R$};
	\draw (1.5,2.1) node {$S_D$};
	\draw (.8,-3.1) node {$S_1$};
  				\draw (4,-3) node {$S_2$};
  				\draw (2.4,-3.5) node {$\underbrace{E}$};
  				\draw (2,.7) node {$(1)$};
  				\draw (4.1,-1.7) node {$(2)$};
  				\draw (-1,-4.5) node {$(iii)$};
  				\end{tikzpicture}
   }
\caption{\textit{Asymptotic solution in neighborhood of $Q'$, which corresponds to the point $Q$ in Figure $1$; $S_1$ and $S_2$ are the sonic lines. Figure $3(i)$ corresponds to a perfect gas case $(\tilde{b}=0)$, whereas $3 (ii)$ and $3 (iii)$ account for the real gas effects with $\tilde{b}=0.3$ and $\tilde{b}=0.6$, respectively, with $\gamma=1.4$.}}
  \label{figure1}
  \end{figure} 
It may be noticed that an increase in $\tilde{b}$ not only causes the sonic lines $S_1$ and $S_2$ to shift along the positive $\tilde{r}$ direction, but it also increases the breadth between them (see Figure $3$). Further, equation (\ref{equ53*}) shows that an increase in $\tilde{b}$ serves to reduce the jump in the derivatives of flow variables across the expansion wave $E$; also, a change in the parabolic configuration in Figure 3 leads us to reinforce our conclusion that the domain of the elliptic region in the neighborhood of the singular point $Q'$ exhibits an increase with an increase in the van der Waals parameter $\tilde{b}$.
%%%%%%%%%%%%%%%%%%%%%%%%%%%%%%%%%%%%%%%%%%%%%%%%%%%%%%%%%%%%%%%%%%%%%%%%%%%%%%%%%%%%%%%%%%%%%%%%%%
\section{Conclusions}
Here, we analyze the problem of a plane shock, reflected and diffracted off a right angled wedge, using asymptotic expansions and explore how the real gas effects influence the configuration of the flow patterns relative to what it would have been in the ideal gas case. In the limit of vanishing van der Waals excluded volume, the ideal gas case presented in the work of Harabetian \cite{eduard1987}, who considered the analogous unsteady wedge-diffraction problem, and Keller \& Hunter \cite{kel}, who analyzed the problem (away from the singular point) using nonlinear ray method, is recovered. We find that the reflected and diffracted regions as well as their boundaries, referred to as wavefronts, are significantly influenced by the real gas effects in the sense that an increase in the van der Waals excluded volume fosters an expansion of the linearized solution domain. As the linearized solution is invalid near the wavefronts due to the directional singularity of type $\sqrt{\zeta-a_0}$, we construct weakly nonlinear solutions near the wavefronts but away from the singular point by following the theory of asymptotic expansions, and match the nonlinear solution with the linearized solution to obtain a uniformly valid asymptotic approximation in the flow field. It is shown that if the diffracted wave is a rarefaction wave, across which there is a continuous transition from the supersonic region to the subsonic region, it gets weakened by the real gas effects and decays slowly as compared to the corresponding ideal gas case. However, if the diffracted wave is a shock, its speed and strength both enhance with an increase in $\tilde{b}$. In order to investigate the nature of the flow in the vicinity of the singular point, where the first order approximation ceases to be valid, asymptotic expansions are constructed which lead to a pair of PDEs, exhibiting a remarkable resemblance with UTSD equations derived in \cite{morawetz, hunter2013}. In the neighborhood of the singular point, the asymptotic location of the reflected shock reinforces our conclusion that the real gas effects cause the boundaries of different flow domains to dilate. Location of the sonic lines in the neighborhood of $Q$, at which the reflected shock starts bending and the equations change type, is shifted in the positive $\tau$ direction. It is shown that the governing asymptotic equation and the auxiliary conditions allow similarity solutions to exist near the singular point; a rarefaction wave solution, satisfying the specific boundary conditions, is obtained. It is concluded that the real gas effects serve to weaken the rarefaction wave and to strengthen the diffracted wave supporting our earlier viewpoint.


\begin{thebibliography}{99}
\bibitem{courant1976} Courant, R., Friedrichs, K. O. (1976). {Supersonic flow and shock waves,} Springer-Verlag, New York.
%\bibitem{glimm} Glimm, J.,  Majda, A. (1991).  Multidimensional hyperbolic problems and computations. {\it IMA volumes in Mathematics and its Applications}, {\bf 29}, Springer-Verlag, NewYork.
\bibitem{glass} Glass, I. I.,  Sislian, J. P. (1994). {Nonstationary flows and shock waves ,} Oxford University Press.
%\bibitem{sharma} Sharma, V. D. (2010). {\it Quasilinear hyperbolic systems and conservation laws,} CRC Press New York.
\bibitem{zheng} Zheng, Y.  (2001). {System of conservation laws: two-dimensional Riemann problems,} Birkhauser, Boston.
\bibitem{zheng1} Zheng, Y.  (2006). Two-dimensional regular shock reflection for the pressure gradient system of conservation laws. {\it Acta Math. Appl. Engl. Ser. }, {\bf 22}, 177-210.
\bibitem{dor} Ben-Dor, G. (2007). {Shock wave reflection phenomena,} 2nd edition, Springer-Verlag, New York.

\bibitem{chen} Chen, G. Q. (2011). Multidimensional conservation laws: overview, problems, and perspective. Nonlinear conservation laws and applications. {\it IMA Vol. Math. Appl.}, {\bf 153}, 23-72.
\bibitem{chang} Chang, T.,  Hsiao, L. (1989). {The Riemann problem and interaction of waves in gas dynamics,} John Wiley and Sons, New York.
\bibitem{keller1951} Keller, J. B.,  Blank, A. A. (1951). Diffraction and reflection of pulses by wedges and corners. {\it Comm. Pure Appl. Math.}, {\bf 4}, 75-94.
\bibitem{kel} Keller, J. B.,  Hunter, J. K. (1984). Weak shock diffraction. {\it Wave Motion}, {\bf 6}, 79-89.
%\bibitem{keller1983} Hunter, J.,  Keller, J. B. (1983). Weakly nonlinear high frequency waves. {\it Comm. Pure Appl. Math.}, {\bf 36}, 547-569.
%\bibitem{majda} Hunter, J. K., Majda, A.,  Rosales, R. (1986). Resonantly interacting, weakly nonlinear hyperbolic waves. II - Several space variables . {\it Std. in Appl. Math.}, {\bf 75}, 187-226.
\bibitem{eduard1987} Harabetian, E. (1987). Diffraction of a weak shock by a wedge. {\it Comm. Pure Appl. Math.}, {\bf XL}, 849-863.
\bibitem{myers} Zahalak, G. I.,  Myers, M. K. (1974). Conical flow near singular rays. {\it J. Fluid Mech}, {\bf 63-3}, 537-561.
\bibitem{hunter88} Hunter, J. K. (1988). Transverse diffraction of nonlinear waves and singular rays. {\it SIAM J. Appl. Math. }, {\bf 48-1}, 187-226.

\bibitem{morawetz} Morawetz, C. S. (1994). Potential theory for regular and Mach reflection of a shock at a wedge. {\it Comm. Pure Appl. Math.}, {\bf 47}, 593-624.
\bibitem{zheng2} Zheng, Y. (1997). Existence of solutions to the transonic pressure gradient equations of the compressible Euler equations in elliptic regions. {\it Comm. Pure Appl. Math.}, {\bf 22}, 1849-1868.
\bibitem{tabak} Rosales, R. R., Tabak, E. G. (1998). Caustics of weak shock waves. {\it Phys. Fluids}, {\bf 10}, 206-222.
\bibitem{Hunter2000} Hunter, J., Brio, M. (2000). Weak shock reflection. {\it J. Fluid Mech.}, {\bf 410}, 235-261.
\bibitem{keyfitz} Canic, S., Keyfitz, B. L., Kim, E. H. (2002). A free boundary problem for a quasilinear degenerate elliptic equation: Regular reflection of a weak shocks. {\it Comm. Pure Appl. Math.}, {\bf 65}, 71-92.
\bibitem{tesdall}  Tesdall, A. M., Hunter, J. K. (2002). Self-Similar solutions for weak shock reflection. {\it SIAM J. Appl. Math.}, {\bf 63}, 42-61.
%\bibitem{thompson} Thompson, P. A. (1971). A fundamental derivative in gasdynamics. {\it Phys. Fluids}, {\bf 14}, 1843-1849.
%\bibitem{cramer} Cramer, M. S., Sen, R. (1987). Exact solutions for sonic shocks in van der Waals gases. {\it Phys. Fluids}, {\bf 30}, 370-385.
%\bibitem{best} Cramer, M. S., Best, L. M. (1991). Steady, isentropic flows of dense gases. {\it Phys. Fluids}, {\bf 3}, 219-226.
%\bibitem{kluwick} Kluwick, A. (1991). {\it Nonlinear waves in real fluids,} Springer-Verlag, Berlin.
\bibitem{Wu1996} Wu, C. C., Roberts, P. H. (1996). Structure and stability of a spherical shock wave in a van der Waals gas. {\it Quart J Mech Appl Math}, {\bf 49}, 501-543.
%\bibitem{shyue} Shyue, K. M. (1999). A fluid-mixture type algorithm for compressible multicomponent flow with van der Waals equation of state. {\it J. comp. Phys.}, {\bf 156}, 43-88.
\bibitem{arora} Arora, R., Sharma, V. D. (2006). Convergence of strong shock in a Van der Waals gas. {\it SIAM J. Appl. Math.}, {\bf 66}, 1825-1837.
\bibitem{manoj}  Pandey, M., Sharma, V. D. (2007). Interaction of a characteristic shock with a weak discontinuity in a non-ideal gas. {\it Wave Motion}, {\bf 44}, 346-354.
\bibitem{manoj1} Pandey, M., Sharma, V. D. (2009). Kinematics of a shock wave of arbitrary strength in a non-ideal gas. {\it  Quart. Appl. Math.}, {\bf 67}, 401-418.
\bibitem{chen1986} Chang, T., Chen, G. Q. (1986). Diffraction of planar shock along a compressive corner. {\it Acta. Math. Sci.}, {\bf 6}, 241-257.
\bibitem{tesdall2} Hunter, J. K., Tesdall, A. M. (2004). {Weak shock reflection in `A celebration of mathematical modeling',} Kluwer academic press, NewYork.
%\bibitem{landau} Landau, L. D. (1945). On shock waves at large distances from their place of origin. {\it Soviet J. Physics}, {\bf 9}, 496-500.
%\bibitem{whith} Whitham, G. B. (1973). {\it Linear and nonlinear waves,} John Wiley and Sons, New York.
\bibitem{hunter2013} Hunter, J. K., Tesdall, A. M. (2012). On the self-similar diffraction of a weak shock into an expansion wavefront. {\it SIAM J. Appl. Math.}, {\bf 72}, 124-143.
%\bibitem{light} Lighthill, M. J. (1949). The diffraction of blast I. {\it Proc. Roy. Soc.}, {\bf A 198}, 454-470.
%\bibitem{ting} Ting, L., Ludloff, H. F. (1952). Aerodynamics of blast. {\it J. Aero. Sci.}, {\bf 19}, 317.
\end{thebibliography}
\end{document}